\documentclass[useAMS,usenatbib]{mn2e}
\usepackage{epsfig}

\title[Metallicity measurements in M and K dwarf stars]{Metallicity measurements
using atomic lines in M and K dwarf stars}
\author[V. M. Woolf and G. Wallerstein]{Vincent M. Woolf$^{1}$\thanks{E-mail:
vmw@astro.washington.edu (VMW); wall@astro.washington.edu (GW)} and George
Wallerstein$^{1}$\footnotemark[1]{} \thanks{Based on observations obtained
with the Apache Point Observatory 3.5-meter telescope, which is owned and
operated by the Astrophysical Research Consortium. This publication makes use
of data products from the Two Micron All Sky Survey, which is a joint project
of the University of Massachusetts and the Infrared Processing and Analysis
Center/California Institute of Technology, funded by the National Aeronautics
and Space Administration and the National Science Foundation.}\\
$^{1}$Astronomy Department, University of Washington, Box 351580, Seattle, WA
98195, USA \\}
\begin{document}

\date{}

\pagerange{\pageref{firstpage}--\pageref{lastpage}} \pubyear{2004}

\maketitle

\label{firstpage}

\begin{abstract}
We report the first survey of chemical abundances in M and K dwarf stars using
atomic absorption lines in high resolution spectra.
We have measured Fe and Ti abundances in
35 M and K dwarf stars using equivalent widths measured from
$\lambda / \Delta \lambda \approx 33\,000$ spectra. Our analysis takes
advantage of recent improvements in model atmospheres of low-temperature dwarf
stars.  The stars have temperatures between 3300 and 4700 K, with most cooler
than 4100 K. They cover an iron abundance range of $\rm -2.44 < [Fe/H] < +0.16$.
Our measurements show [Ti/Fe] decreasing with increasing [Fe/H], a trend
similar to that measured for warmer stars where abundance analysis techniques
have been tested more thoroughly.  This study is a step toward the observational
calibration of procedures to estimate the metallicity of low-mass dwarf stars
using photometric and low-resolution spectral indices.

\end{abstract}


\section{Introduction}

Low mass (M dwarf and cooler) main sequence stars are by far the most numerous
stars in the Galaxy and make up most of its baryonic mass. However, there have
been few detailed chemical abundance studies of these stars with spectra of
sufficient resolution for atomic absorption lines to be measured individually.
None of these
included more than a few stars.  This was largely a result of their low
intrinsic luminosity and the molecular lines present in their spectra. Their
faintness meant that few of these stars were bright enough for high signal
to noise, high
resolution spectra to be measured easily without a large telescope.
The molecular bands present in their spectra
complicated the calculation of stellar model atmospheres and cause line blends
which make it difficult to measure atomic line strengths in large regions of
the visible spectrum.  There have been abundance studies which used photometry
or which fitted broad molecular features in low-resolution spectra of M and K
dwarfs, but their
results are not as certain as they would be if they were calibrated with
stars for which more precise abundances were based on higher-resolution spectra.
There have also been studies which used synthetic spectrum fitting of low
resolution spectra to measure metallicities of low mass dwarfs. These have given
rough metallicity estimates, but have not produced elemental abundances with the
precision which is possible using higher resolution spectra.

Recent advances in model atmospheres of low mass dwarf stars, largely as a
result of the improved treatment of molecular opacity, have provided
the opportunity to determine abundances of M and K dwarf stars through
analysis of their atomic spectral lines.  Our study of chemical abundances in
Kapteyn's Star (HD 33793) \citep{ww04} showed us
that it is possible to locate spectral regions in these stars where molecular
bands are not present or are sufficiently weak for accurate equivalent
widths of atomic lines to be measured. 

The survey of abundance estimates we report here will provide some of the data
necessary to calibrate metallicity indices for cool dwarf stars, for example
the TiO and CaH indices of \citet{rhg95} and \citet{hgr96}.  With 
calibrated indices it should be possible to estimate the metallicity of
a much larger number of cool dwarfs, including those too distant and faint
to study using high resolution spectra.  When a large volume-limited sample
is available, it will be possible to determine if the `G dwarf problem'
\citep{vdb62, at76} continues to lower-mass main sequence stars so that
low-metallicity M dwarfs are more scarce than expected in our Galaxy.

\section{Observations and reduction}

We selected stars for observation to cover a large range of metallicity: low
metallicity stars are overrepresented in our sample compared to the actual
number in the solar neighborhood.  We increased the number of low metallicity
stars by choosing to observe stars with high radial velocities, which increased
the likelihood of observing halo stars, and by observing stars with TiO5
indices \citep{g97} which indicate weak TiO band strengths.

Our spectrum of Kapteyn's Star was measured with the echelle
spectrograph of the 4.0-m Victor M. Blanco Telescope at the Cerro Tololo
Inter-American Observatory as described in \citet{ww04}.  The 34 other stars
were observed using the echelle spectrograph of the Apache Point Observatory
(APO) 3.5-meter telescope.

The spectra were reduced using standard IRAF routines to subtract the bias,
divide by flat field spectra, reduce the echelle orders to one dimensional
spectra, and apply ThAr lamp spectrum wavelength calibration. The spectrum of
a hot, high $v \sin i$ star was used to correct for telluric absorption lines
where appropriate and possible.

The spectral resolution of the APO spectra is about  $\lambda / \Delta \lambda 
\approx 33\,000$, as measured using the ThAr comparison lines. There
are no gaps in wavelength between echelle orders. The usable spectrum covers
the range from about 9800 \AA\ to where the measured signal from these
red stars drops off in the blue, normally below 5000 \AA.  For most stars, the
signal to noise ratio in the region of the lines we used for the analysis was
at least 100 per pixel.  It was much higher for the brighter
stars ($V \la 11$) in our sample. For the faintest star, LHS 364, we were
only able to achieve a signal to noise ratio of about 50 with the observation
time available.

\section{Analysis}

\subsection{Stellar parameters and model}

As we found in our analysis of Kapteyn's Star, the chemical
abundances we derive for these low-temperature dwarf stars depends strongly on
the metallicity of the model atmosphere used.  For example, changing the
metallicity of the model atmosphere with $T_{\rm eff} = 3500$~K and
$\log g = 5.0$ by $\pm 0.5$ dex can changes the Fe abundance derived using the
equivalent widths measured for Kapteyn's Star by $\pm 0.3$ dex. 
Determining the physical
parameters to use for the model atmospheres is therefore an iterative process.
Fortunately, the iteration converges so it is possible to find the value
where the metallicity derived equals the metallicity of the model atmosphere
used in the analysis.

We integrated the flux in the
$V$, $H$, and $K_{\rm S}$ filters for a grid of synthetic spectra released
with the NextGen models (Hauschildt, Allard, \& Baron 1999)
to produce theoretical colour-temperature relation estimates.
We obtained $H$ and $K_{\rm S}$ magnitudes from the 2MASS point source catalog
\citep{c03} and $V$ from \citet{mmh97} and used
these to find $V - K_{\rm S}$ and $V - H$ temperatures for our stars. We used
the average of the two as $T_{\rm eff}$ in our models.
We note that while there are few other published determinations of temperatures
in the stars we observed, we were happy to see that the temperature we
derive for HD~88230, $T_{\rm eff} = 3970 \pm 220$~K, is in good agreement with
the temperature derived by \citet{rm04}, $T_{\rm eff} = 3962 \pm 63$~K,
using its bolometric flux and its measured angular diameter.

For stars where parallax data was
available we calculated absolute $H$ and $K_{\rm S}$ magnitudes.  We used these
to estimate the masses using the theoretical mass-luminosity relations plotted
by \citet{sdf03}. We calculated the bolometric correction ${\rm BC}_K$ from
the ${\rm BC}_V$ and $V-K$ NextGen colours \citep{hab99} and used 
parallax, $K_{\rm S}$ and ${\rm BC}_K$ to derive $\rm M_{bol}$.  We then used
$\rm M_{bol}$, mass, and $T_{\rm eff}$ to calculate $\log g$:
\[
\log g = \log M + 4\log(T_{\rm eff}/5770) + 0.4({\rm M_{bol}} - 4.65) + 4.44
\]
where we have used $T_{\rm eff\odot} = 5770$~K, $\rm M_{bol\odot} = 4.65$,
$\log g_\odot = 4.44$, and $M$ is in solar masses. 
For stars where parallax measurements were unavailable we assumed
$\log g = 5.0 \pm 0.5$. Gravity does not have an effect on the derived
chemical abundances as large as the effects of temperature or model
metallicity: accepting this large gravity uncertainty does not produce a
large uncertainty in the abundances.  The parallaxes and magnitudes
used to derive the stellar physical parameters are listed in Table~\ref{T1}.
Spectral types are included in the table to give a rough idea of how low
resolution spectra of the stars appear: the temperatures and gravities
derived for the stars and reported in Table~\ref{T2} are more
physically meaningful.

\begin{table*}
\begin{minipage}[l]{110mm}
\caption{M and K dwarf magnitudes and parallaxes}\label{T1}
\begin{tabular}{lllrcrcrcrrc}
\hline
Star &Alternate & Spectral& V & $\pm$& K$\rm _s$ & $\pm$& H & $\pm$& $\pi$ &
$\pm$& $\pi$ source\footnote{H: ESA (1997), Y: van Altena, Lee, \&
Hoffleit (1995), N: Harrington \& Dahn (1980), we assume
that HD 97101B has the same parallax as HD 97101A} \\
 & name & type\footnote{taken from Gizis (1997), SIMBAD, Lee (1984), and Luyten
(1979)}     &&&&  &          &   & mas & mas \\
\hline
HD 33793&GJ 191 & sdM1.0& 8.85& 0.03&  5.05& 0.02&  5.32& 0.03& 255.1&  0.9&H \\
HD 36395&GJ 205 & M1.5&  7.96& 0.01&  4.04& 0.26&  4.15& 0.21& 175.7&  1.2&H \\
HD 88230&GJ 380 & K5 &  6.60& 0.02&  2.96& 0.29&  3.30& 0.26& 205.2&  0.8&H \\
HD 95735&GJ 411 & M2V &  7.49& 0.02&  3.25& 0.31&  3.64& 0.20& 392.5&  0.9&H \\
HD 97101B&GJ 414B&M1.5&  9.98& 0.04&  5.73& 0.02&  5.98& 0.02&  83.8&  1.1&H \\
HD 119850&GJ 526 &M1.5 &  8.46& 0.01&  4.42& 0.02&  4.78& 0.21& 184.1&  1.3&H \\
HD 178126&G 22-15& K5V&  9.23& 0.02&  6.47& 0.02&  6.57& 0.02&  41.2&  1.3&H \\
HD 199305&GJ 809 & M0.5&  8.54& 0.04&  4.62& 0.02&  4.92& 0.06& 142.0&  0.8&H \\
HD 217987&GJ 887 & M0.5&  7.35& 0.02&  3.46& 0.20&  3.61& 0.23& 303.9&  0.9&H \\
LHS 12&HIP 9560 &M0.5& 12.26& 0.04&  8.68& 0.02&  8.90& 0.03&  36.1&  4.3&H \\
LHS 38&GJ 412A & M0.5&  8.75& 0.04&  4.77& 0.02&  5.00& 0.02& 206.9&  1.2&H \\
LHS 42&GJ 9371 &sdM0.0& 12.20& 0.03&  8.67& 0.02&  8.90& 0.02&  44.3&  2.8&H \\
LHS 104&G 30-48&esdK7& 13.74& 0.02& 10.41& 0.02& 10.57& 0.03&  19.3&  3.0&Y \\
LHS 170&HIP 15234&sdK& 10.68& 0.01&  7.60& 0.02&  7.77& 0.03&  30.2&  2.4&H \\
LHS 173&HIP 16209 &sdK7& 11.11& 0.01&  7.79& 0.02&  7.97& 0.02&  39.2&  2.5&H \\
LHS 174&G 38-2&sdM0.5& 12.75& 0.01&  9.14& 0.02&  9.35& 0.02&  22.6&  7.4&Y \\
LHS 182&GJ 1064D&esdM0.0&13.90& 0.02& 10.52& 0.02& 10.67& 0.03&  23.1&  2.8&Y \\
LHS 236&G 251-44& sdK7& 13.10& 0.01&  9.85& 0.02& 10.00& 0.02&  18.2&  2.9&Y \\
LHS 343&G 61-21&sdK& 13.82& 0.02& 10.66& 0.02& 10.86& 0.02&  18.6&  3.7&Y \\
LHS 364&GJ 3825&esdM1.5&14.55&0.03& 10.86& 0.01& 11.01& 0.02&  36.1&  3.2&Y,N \\
LHS 450&GJ 687 &M3&  9.15& 0.03&  4.55& 0.02&  4.77& 0.03& 220.9&  0.9&H \\
LHS 467&HIP 91668&esdK7& 12.21& 0.03&  8.78& 0.02&  9.00& 0.02&  26.0&  3.6&H \\
LHS 1138&G 60-18 & G - K&13.29& 0.01& 10.76& 0.02& 10.92& 0.02&   9.5&  4.6&Y \\
LHS 1482&L 586-41& K& 13.96&0.03& 10.82& 0.02& 10.98& 0.02\\
LHS 1819&HIP 28940& K4& 10.88& 0.02&  8.29& 0.03&  8.37& 0.05&  17.0&  2.6&H \\
LHS 1841&L 812-11& K& 13.18&0.03& 10.39& 0.02& 10.51& 0.02&  17.5&  3.3&Y \\
LHS 2161&G 48-21 & K5& 11.58& 0.01&  8.75& 0.02&  8.83& 0.05\\
LHS 2463&G 10-53 &K7& 12.48& 0.01&  9.83& 0.02&  9.99& 0.03\\
LHS 2715&GJ 506.1& sdK& 10.84& 0.02&  8.17& 0.02&  8.31& 0.03&  27.9&  2.5&H \\
LHS 2938&HIP 71122& K7& 10.67& 0.02&  7.76& 0.02&  7.95& 0.05&  19.0&  2.0&H \\
LHS 3084&G 15-26& sdK& 13.43& 0.03&  9.78& 0.02&  9.99& 0.03&  19.1&  2.9&Y \\
LHS 3356&GJ 701 &M1&  9.37& 0.03&  5.31& 0.02&  5.57& 0.04& 128.3&  1.4&H \\
LHS 5337&G 21-12& M0& 11.15& 0.03&  7.47& 0.02&  7.66& 0.05&  34.5&  3.3&H \\
G 39-36&G 86-10& & 12.36& 0.01&  9.54& 0.02&  9.76& 0.03\\
HIP 27928&GJ 9192& K4& 10.70& 0.02&  7.76& 0.02&  7.88& 0.03&  26.1&  2.1&H \\
\hline
\end{tabular}
\end{minipage}
\end{table*}

We obtained from P. Hauschildt (private communication) an updated NextGen model
atmosphere grid which improves on the most recent public release \citep{hab99}
by including the improved TiO and $\rm H_2O$ line lists described in
\citet{ahs00} in its calculation.  We used this to
create model atmospheres interpolated in $\log g$ and $T_{\rm eff}$ for
each star.

For each star we began by assuming a metallicity
of $\rm [M/H] = -1.0$.\footnote{We use the standard notation
$\rm [X] \equiv \log_{10}(X)_{star} - \log_{10}(X)_\odot .$ }
The calculated Fe and Ti abundances were used to estimate the model atmosphere
metallicity to be used in the next iteration.  The temperature
and gravity calculated for the stars also depend on the assumed metallicity,
so these also varied during the iteration process.  This procedure was repeated
until the metallicity derived from the abundances equalled that of the
model atmosphere used to calculated them. We note that in this case
we define the ``metallicity'' value by the effect of metals in the
stellar atmosphere, primarily through their effect on continuous opacity, not 
by the total concentration of all elements heavier than He.

Model metallicity corrections due to non-solar [$\alpha$/Fe] abundances were
estimated using the LTE stellar analysis program MOOG \citep{s73} output. 
The average Ti abundance at a star's [Fe/H] was
used as a proxy for the star's $\alpha$ element enhancement.  MOOG provides
partial pressures of requested species at the different layers of the model
atmosphere, so by finding, for example, the Mg I and Mg II partial pressures we
were able to estimate the fraction of Mg which is ionized at the model layer
where the reference opacity (at 1.2 $\mu$m) is about 0.1,
approximately where the lines in which we are interested
are formed.  The ionization fractions of
Na, Mg, Ca, Al, and Fe, the major electron donors, were found for each model
in the grid, and thus the fraction of free electrons provided by the $\alpha$
elements Mg and Ca were estimated.  The model metallicity then used for a star
was adjusted to account for how the free electron density was affected by the
non-solar [$\alpha$/Fe] abundances.  The majority of the continuous
opacity in the line-forming regions in the atmospheres of these cool dwarfs
is provided by the $\rm H^-$ ion, and is thus proportional to the number of free
electrons.

For Kapteyn's Star (HD 33793) we used the temperature and gravity derived by 
\citet{skf03} using radius measurements the Very Large Telescope Interferometer,
rather than using our photometry and parallax procedure.

The microturbulence parameter was estimated by requiring that there be no 
slope in Ti abundance vs equivalent width. At the temperatures of these stars,
Ti lines are more common than the Fe lines which are normally used for this
purpose in warmer stars.

\subsection{Chemical abundances}

For this survey we have chosen to report our results only for Fe and Ti since
there are many more reliable lines of those species than for any other
elemenst.  We measured equivalent widths
of Fe~{\sc i} and Ti~{\sc i} lines in the spectra
using the SPLOT routine of IRAF.  We used lines in the cleanest spectral regions
available, where the effects of molecular bands and telluric lines were
minimized. Because the molecular line strengths depend on stellar
temperature and metallicity, we were more successful in finding clean lines for
some stars than in others. 
We accepted partly blended lines for analysis if the blending
occurred far enough from the line centre that we were able to use the
deblending feature of SPLOT to remove the effect the adjacent line(s).

We did not measure equivalent widths in the spectra of four of the stars we
observed for this project.  The spectrum of Barnard's Star (LHS 57) had few
regions free of molecular bands so that very few unblended atomic lines could be
found.  LHS 537 and LHS 1718 appear to be double-lined spectroscopic binaries,
the analysis of which is beyond the scope of this project.
G 30-2 has the broad lines of a fast rotating star, which meant that we
were unable to find any unblended lines in its spectrum.

The line data were compiled using the Vienna Atomic Line Database (VALD)
\citep{vald} and the Kurucz atomic spectral line database \citep{kb95}.  The
line data and equivalent widths measured for each star are listed in
Table~\ref{T2}.  The original sources of the atomic data for the
lines are included in Table~1 of \citet{ww04}.

\begin{table}
\begin{minipage}[l]{70mm}
\caption{Fe and Ti line data}\label{T2}
\begin{tabular}{lccrc}
\hline
wavelength & $\chi$ & $\log gf$ & EW & $\log N$  \\
(\AA )     &  (eV)  &           & (m\AA ) \\
\hline

 {\bf HD 33793 } \\
 Fe {\sc i} \\
 8047.62 & 0.86 & -4.656 &  91.0 &  6.42 \\
 8075.15 & 0.91 & -5.062 &  43.0 &  6.36 \\
 8327.06 & 2.20 & -1.525 & 338.0 &  6.53 \\
 8387.77 & 2.18 & -1.493 & 355.0 &  6.52 \\
 8514.07 & 2.20 & -2.229 & 160.0 &  6.47 \\
 8515.11 & 3.02 & -2.073 &  44.0 &  6.52 \\
 8582.26 & 2.99 & -2.133 &  38.0 &  6.44 \\
 8611.80 & 2.85 & -1.926 &  81.0 &  6.51 \\
 8621.60 & 2.95 & -2.321 &  30.5 &  6.45 \\
 8674.75 & 2.83 & -1.800 &  95.0 &  6.48 \\
\hline
\end{tabular}
The full table is available in the electronic version.
\end{minipage}
\end{table}

We used MOOG to calculate the abundances from equivalent widths.
As discussed previously, determining the stellar chemical abundances 
is an iterative procedure for these low mass dwarfs which
continued until the metallicity calculated using the Fe and Ti abundances
equalled the model atmosphere metallicity.

\begin{figure}
\epsfig{file=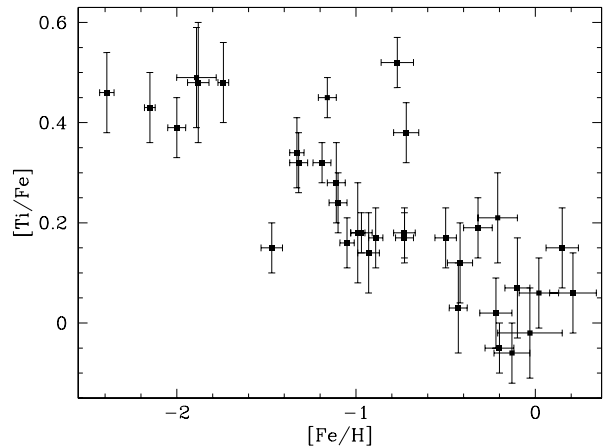, width=6.2cm, angle=270}
\caption{[Ti/Fe] vs [Fe/H] for 35 M and K dwarf stars.}
\label{F1}
\end{figure}

The Fe and Ti abundances we found and the stellar parameters, including
$\alpha$-element weighted metallicity used in the model atmospheres,
are listed in Table~\ref{T3}. The
uncertainties listed include the effects of the uncertainties of the
parallax and photometry data used to derive stellar parameters and the
statistical scatter of abundances calculated from different lines in the same
star. The ratio [Ti/Fe] is plotted against [Fe/H] in Figure~\ref{F1}.

We measured equivalent widths of our Fe {\sc i} and Ti {\sc i} lines in the
solar spectrum
\citep{kfb84} where they were not obscured by blending.  When we calculated the
abundances using the solar equivalent widths and a Kurucz model atmosphere
with the solar values $T_{\rm eff} = 5777$~K, $\log g = 4.44$, and
$\rm \xi = 1.15\ km\ s^{-1}$, we found $A({\rm Fe}) = 7.48$ and
$A({\rm Ti}) = 4.99$.  Altering the $gf$ values of the lines to produce
solar $gf$ values would result in very little change in our results, as we
have assumed the solar abundances are $A({\rm Fe}) = 7.45$ and
$A({\rm Ti}) = 5.02$ \citep{l03}.

\begin{table*}
\begin{minipage}[l]{150mm}
\caption{M and K dwarf parameters and abundances}\label{T3}
\begin{tabular}{lccllcrrcrcrc}
\hline
Star & T$_{\rm eff}$ & $\pm$& $\log g$\footnote{we used $\log g = 5.0 \pm 0.5$
for the four stars where parallax was unavailable.}  & $\pm$ &
$\xi$&[M/H]& [Fe/H] & $\pm$ & [Ti/H] & $\pm$ & [Ti/Fe]\footnote{we use
$A{\rm (Fe)}_\odot = 7.45$, $A{\rm (Ti)}_\odot = 5.02$ } & $\pm$\\
        && K & && km s$^{-1}$ &\\
\hline
HD 33793& 3570&160&  4.96& 0.13& 2.00& $-$0.86& $-$0.99& 0.04& $-$0.81& 0.09&  0.18& 0.10 \\
HD 36395&3760&140&  4.71& 0.20& 1.00&  0.15&  0.21& 0.13&  0.27& 0.13&  0.06& 0.08 \\
HD 88230&3970&220&  4.51& 0.22& 1.00& $-$0.05& $-$0.03& 0.18& $-$0.05& 0.13& $-$0.02& 0.09 \\
HD 95735&3510&150&  4.82& 0.24& 1.00& $-$0.40& $-$0.42& 0.07& $-$0.30& 0.09&  0.12& 0.08 \\
HD 97191B&3610& 40&  4.65& 0.05& 1.50& $-$0.01& 0.02& 0.11&  0.08& 0.11&  0.06& 0.07 \\
HD 119850& 3650& 40&  4.79& 0.05& 0.50& $-$0.12& $-$0.10& 0.07& $-$0.03& 0.08&  0.07& 0.10 \\
HD 178126& 4530& 30&  4.57& 0.05& 1.00& $-$0.61& $-$0.72& 0.07& $-$0.34& 0.09&  0.38& 0.06 \\
HD 199305& 3720& 50&  4.67& 0.05& 2.00& $-$0.14& $-$0.13& 0.10& $-$0.19& 0.11& $-$0.06& 0.06 \\
HD 217987& 3680&130&  4.88& 0.16& 1.00& $-$0.22& $-$0.22& 0.09& $-$0.20& 0.08&  0.02& 0.07 \\
LHS 12&3830& 40&  4.95& 0.14& 0.50& $-$0.77& $-$0.89& 0.04& $-$0.72& 0.05&  0.17& 0.06 \\
LHS 38&3600& 30&  4.90& 0.04& 1.00& $-$0.40& $-$0.43& 0.05& $-$0.40& 0.09&  0.03& 0.09 \\
LHS 42&3860& 30&  5.05& 0.09& 0.50& $-$0.90& $-$1.05& 0.04& $-$0.89& 0.04&  0.16& 0.05 \\
LHS 104&3970& 30&  5.07& 0.17& 1.00& $-$1.09& $-$1.33& 0.04& $-$0.99& 0.07&  0.34& 0.07 \\
LHS 170&4230& 30&  4.64& 0.10& 1.00& $-$0.81& $-$0.97& 0.06& $-$0.79& 0.05&  0.18& 0.04 \\
LHS 173&4000& 20&  4.75& 0.08& 1.00& $-$0.98& $-$1.19& 0.05& $-$0.87& 0.05&  0.32& 0.04 \\
LHS 174&3790& 20&  4.78& 0.31& 1.00& $-$0.95& $-$1.11& 0.05& $-$0.83& 0.07&  0.28& 0.08 \\
LHS 182&3870& 30&  5.09& 0.14& 1.00& $-$1.88& $-$2.15& 0.03& $-$1.72& 0.08&  0.43& 0.07 \\
LHS 236&4040& 20&  4.92& 0.16& 1.50& $-$1.07& $-$1.32& 0.05& $-$1.00& 0.06&  0.32& 0.06 \\
LHS 343&4110& 30&  5.10& 0.21& 1.00& $-$1.45& $-$1.74& 0.03& $-$1.26& 0.07&  0.48& 0.08 \\
LHS 364&3720& 30&  5.44& 0.12& 0.50& $-$0.82& $-$0.93& 0.06& $-$0.79& 0.07&  0.14& 0.08 \\
LHS 450&3340& 20&  4.82& 0.03& 1.00&  0.10&  0.15& 0.09&  0.30& 0.11&  0.15& 0.08 \\
LHS 467&3930& 40&  4.83& 0.15& 1.25& $-$0.93& $-$1.10& 0.05& $-$0.86& 0.07&  0.24& 0.06 \\
LHS 1138&4620& 30&  4.86& 0.45& 1.00& $-$2.17& $-$2.39& 0.04& $-$1.93& 0.08&  0.46& 0.08 \\
LHS 1482&4100& 40&  5.0& 0.5& 2.00& $-$1.59& $-$1.88& 0.06& $-$1.40& 0.09&  0.48& 0.12 \\
LHS 1819&4670& 40&  4.54& 0.17& 1.50& $-$0.65& $-$0.77& 0.09& $-$0.25& 0.10&  0.52& 0.05 \\
LHS 1841&4440& 40&  5.13& 0.20& 1.50& $-$1.18& $-$1.47& 0.06& $-$1.32& 0.07&  0.15& 0.05 \\
LHS 2161&4500& 30&  5.0& 0.5& 1.00& $-$0.29& $-$0.32& 0.08& $-$0.13& 0.09&  0.19& 0.06 \\
LHS 2463&4540& 30&  5.0& 0.5& 1.50& $-$1.62& $-$1.89& 0.11& $-$1.40& 0.09&  0.49& 0.10 \\
LHS 2715&4590& 40&  4.85& 0.11& 1.00& $-$0.95& $-$1.16& 0.05& $-$0.71& 0.06&  0.45& 0.04 \\
LHS 2938&4490& 50&  4.44& 0.13& 1.50& $-$0.20& $-$0.21& 0.11&  0.00& 0.16&  0.21& 0.09 \\
LHS 3084&3780& 30&  4.88& 0.16& 1.00& $-$0.64& $-$0.73& 0.05& $-$0.56& 0.05&  0.17& 0.05 \\
LHS 3356&3630& 30&  4.79& 0.04& 1.50& $-$0.20& $-$0.20& 0.08& $-$0.25& 0.09& $-$0.05& 0.05 \\
LHS 5337&3780& 40&  4.59& 0.12& 0.50& $-$0.45& $-$0.50& 0.06& $-$0.33& 0.05&  0.17& 0.06 \\
G 39-36& 4400& 50&  5.0& 0.5& 1.50& $-$1.71& $-$2.00& 0.05& $-$1.61& 0.07&  0.39& 0.06 \\
HIP 27928& 4370& 30&  4.64& 0.10& 1.00& $-$0.62& $-$0.73& 0.06& $-$0.55& 0.08&  0.18& 0.05 \\
\hline
\end{tabular}
\end{minipage}
\end{table*}

\begin{figure}
\epsfig{file=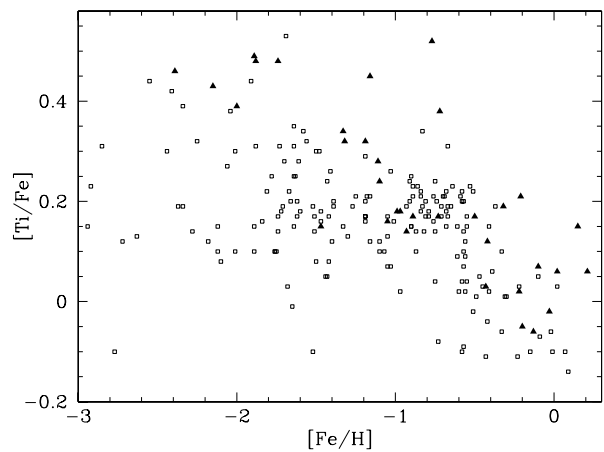, width=6.2cm, angle=270}
\caption{[Ti/Fe] vs [Fe/H]. Triangles represent our M and K dwarf stars. Small
squares represent the warmer dwarfs and subgiants catalogued by Fulbright
(2000).}
\label{F2}
\end{figure}

\section{Discussion}

We have calculated the Fe and Ti abundances of 35 M and K dwarf stars using
atomic line equivalent widths measured using spectra with high spectral
resolution and high signal-to-noise ratios.
We believe this more than triples the number of dwarf stars in the temperature
range of our sample for which chemical abundances have been measured using
atomic lines.  Our abundance estimates should be more reliable than those of
previous studies because we have used updated model atmospheres which include
molecular opacity data which is more complete than what was available in any
previous models. 

The stars we studied have temperatures between 3300 and 4700 K, with a median
temperature of 3950 K.  They have [Fe/H] abundances between $-2.4$ and $+0.2$
The abundance ratio [Ti/Fe] decreases with increasing [Fe/H], showing a
trend similar to what has been observed for Ti and other $\alpha$-elements in
warmer stars \citep{w62}.

The previous studies using atomic lines and high resolution spectra to
measure abundances in dwarf stars in the
temperature range of our stars are sparse and included few stars.  \citet{m76} studied Kapteyn's Star.  He reported
$\rm [Fe/H] = -0.5 \pm 0.3$, which is 0.5 dex larger than our result.
\citet{s94} studied six M and K dwarfs using two Fe~{\sc i} lines.  One star,
HD 95735 (Gl 411), was also included in our study.  The Fe abundance we derived
for the star is 0.42 or 0.57 dex smaller, depending on which gravity Savanov
used.

Several researchers have estimated metallicities of low mass dwarfs by fitting
synthetic spectra to low resolution observed spectra,
$400 < \lambda / \Delta \lambda < 6000$, which included atomic and molecular
lines \citep{j95,jlah96,viti97,leg02}. 
The spectra were for the most part in the
near infrared.  The metallicities derived were not determined as precisely as
is possible using higher resolution spectra, but the method shows what analysis
is possible for fainter stars where low resolution spectra are all that is
available.  

The trend we see in [Ti/Fe] vs [Fe/H] is similar to that seen for warmer stars
in the Galaxy. Figure~\ref{F2} compares our data to
the abundances found by \citet{f00}, who observed field halo and disk stars,
most of which are dwarf stars between 5000 and 6500 K.  We have made the
correction necessary to account for the different assumed solar abundances.  
Our abundance values do not appear out of place among the larger set, although
our average [Ti/Fe] is higher by about 0.1 dex at most [Fe/H] values. We
see a similar relation between our data and that compiled by \citet{gcc03}. The
similar [Ti/Fe] trend was expected since stars of all masses are presumably
made from clouds with the same compositions.

Because we selected stars for observation to cover a wide range of metallicity,
rather than to statistically represent the low mass dwarfs in a given volume,
our results say nothing about the relative numbers of stars with different
metallicities.

One goal of this work will be to find a combination of photometric and
low-resolution spectral indices which can be used to determine the metallicity
of these low mass dwarfs, and to calibrate this method using our observationally
determined metallicities.  Because the molecular band strengths measured by the
spectral indices depend on both the chemical composition and the temperature of
a star, the method will need to be sensitive to both properties. We plan to 
obtain low resolution spectra of the stars in our list for which TiO and CaH
indices have not been measured.

By compiling a statistically significant sample of observationally calibrated
metallicity estimates for low mass dwarfs it will be possible to determine the
metallicity distribution of these stars in the Galaxy.  This will provide
observational evidence of whether there is a K and M dwarf problem in our
Galaxy similar to the G dwarf problem, where low-metallicity stars are more
scarce than predicted by Galactic star formation and chemical evolution models.
The answer to this question, positive or negative, will provide important
constraints for models of the chemical enrichment of the Galaxy.

\section*{Acknowledgments}

We thank David Yong for help getting NextGen models to work in MOOG,
Peter Hauschildt for providing pre-release NextGen atmospheres for our use,
and  Suzanne Hawley for helpful discussions about low mass subdwarfs.
This research has made use of the SIMBAD database, operated at CDS, Strasbourg,
France. This research has made use of NASA's Astrophysics Data System
Bibliographic Services.
The authors gratefully acknowledge the financial support of the
Kennilworth Fund of the New York Community Trust.

\label{lastpage}
\end{document}